\documentclass[aps,prl,twocolumn,superscriptaddress,longbibliography]{revtex4-2}
\usepackage{amsmath}
\usepackage{hyperref}
\hypersetup{
	colorlinks=true,
	linkcolor=blue,
	filecolor=blue,
	citecolor = black,      
	urlcolor=cyan,
}

\usepackage{graphicx}
\usepackage{siunitx}
\graphicspath{{./Figures/}}

\usepackage{xargs}                      
\usepackage[pdftex,dvipsnames]{xcolor}
\usepackage{color}

\newcommand{\magn}[1]{\left|#1\right|}

\newcommand{\bra}[1]{\left\langle#1\right|}
\newcommand{\ket}[1]{\left|#1\right\rangle}

\newcommand{\tr}[1]{\mathrm{Tr}\left[ #1 \right]}
\newcommand{\trover}[2]{\mathrm{Tr}_{#1}\left[ #2 \right]}

\newcommand{\EPR}{\mathrm{EPR}}
\newcommand{\od}{1D }
\newcommand{\se}{S^{(2)}}
\newcommand{\chr}{\mathcal{E}_{\textrm{s}}}
\newcommand{\czo}{\mathrm{CZ}_{\mathrm{(odd)}}}
\newcommand{\cze}{\mathrm{CZ}_{\mathrm{(even)}}}
\newcommand{\gft}{\mathrm{GF}(2)}

\begin{document}

\date{\today}
\title{Deterministic Fast Scrambling with Neutral Atom Arrays}
\author{Tomohiro Hashizume}
\thanks{T.H. and G.S.B. contributed equally to this work.}
\affiliation{Department of Physics and SUPA, University of Strathclyde, Glasgow G4 0NG, United Kingdom}
\author{Gregory S.~Bentsen}
\thanks{T.H. and G.S.B. contributed equally to this work.}
\affiliation{Martin A. Fisher School of Physics, Brandeis University, Waltham, Massachusetts 02465, USA}
\author{Sebastian~Weber}
\affiliation{Institute for Theoretical Physics III and Center for Integrated Quantum Science and Technology, University of Stuttgart, 70550 Stuttgart, Germany}
\author{Andrew J.~Daley}
\affiliation{Department of Physics and SUPA, University of Strathclyde, Glasgow G4 0NG, United Kingdom}

\begin{abstract}
Fast scramblers are dynamical quantum systems that produce many-body entanglement on a timescale that grows logarithmically with the system size $N$. We propose and investigate a family of deterministic, fast scrambling quantum circuits realizable in near-term experiments with arrays of neutral atoms.
We show that three experimental tools -- nearest-neighbor Rydberg interactions, global single-qubit rotations, and shuffling operations facilitated by an auxiliary tweezer array -- are sufficient to generate nonlocal interaction graphs capable of scrambling quantum information using only $\mathcal{O}(\log N)$ parallel applications of nearest-neighbor gates. These tools enable direct experimental access to fast scrambling dynamics in a highly controlled and programmable way and can be harnessed to produce highly entangled states with varied applications.

\end{abstract}

\maketitle



%

Quantum information scrambling describes a process in which initially localized quantum information is delocalized by the dynamics of a many-body system and encoded into a many-body entangled state \cite{Page1993,sekino2008,lashkari2011towards,hosur2016chaos}, thereby effectively hiding the information from local observers.
This process cannot occur instantaneously: the fast scrambling conjecture states that scrambling can develop on timescales no shorter than $t_* \gtrsim \log N$, which scale logarithmically with the system size $N$. Systems that saturate this conjectured bound on the scrambling time $t_*$ are known as ``fast scramblers'' \cite{sekino2008}. Fast scrambling dynamics can rapidly generate Page-scrambled states, 
pure quantum states of a many-body system whose reduced density matrix $\rho_A$ is maximally mixed for almost all subsystems $A$ of size $\magn{A} <  N/2$ \cite{Page1993,sekino2008,lashkari2011towards}.
Prototypical models for fast scrambling \cite{ye1993solvable,lashkari2011towards,kitaev2015,maldacena2016remarks,bentsen2019fast,piroli2020random}, inspired by the study of quantum information in black holes \cite{Page1993,hayden2007black,sekino2008}, often feature randomness and long-range couplings as key ingredients,
although some recent deterministic models have been proposed with sparse or all-to-all coupling graphs with varying weights \cite{bentsen2019treelike,Belyansky2020,li2020fast}.
\begin{figure}[t!]
    \includegraphics[width=\columnwidth]{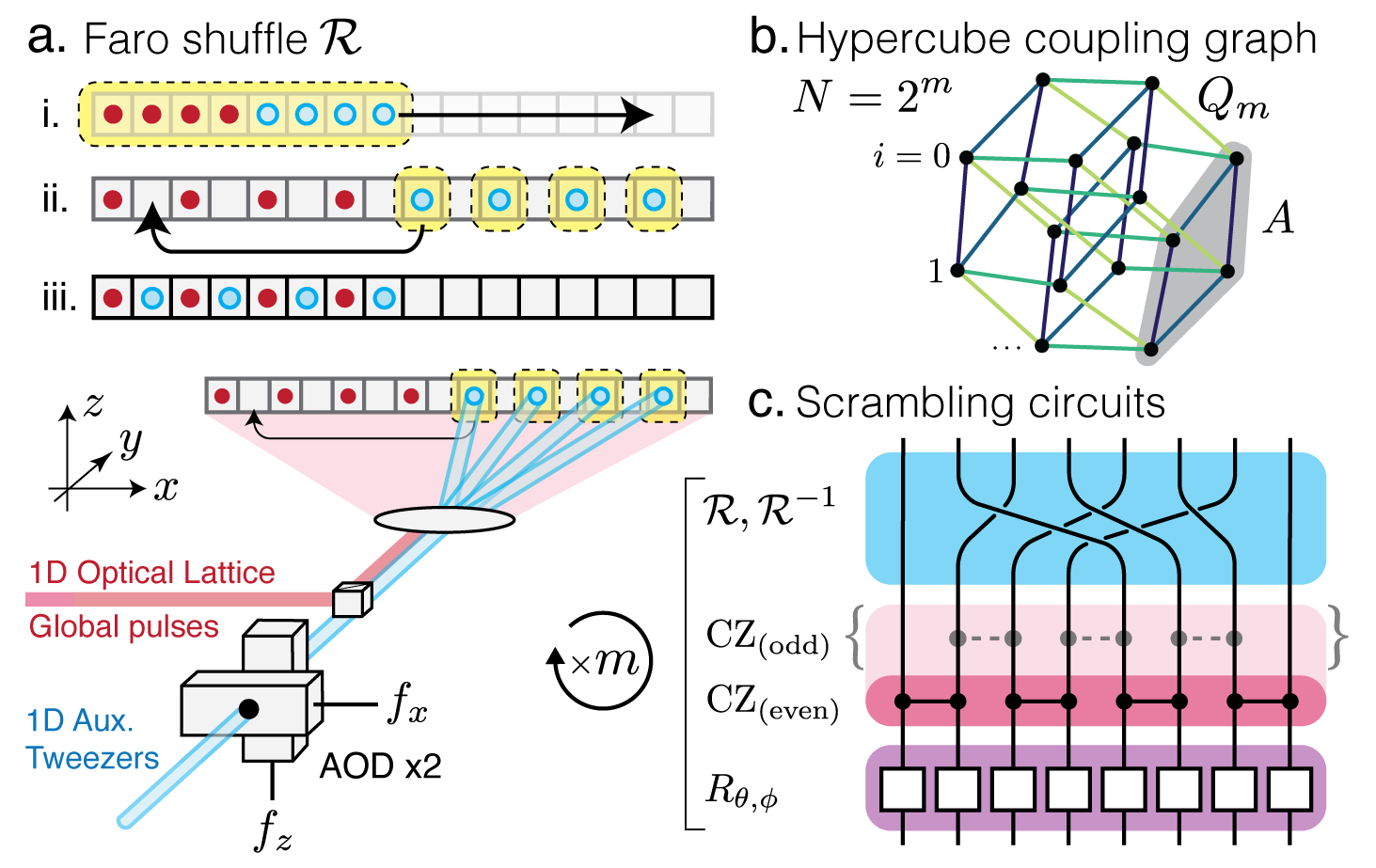}
    \caption{
    Fast scrambling via quasi-\od shuffling. (a) Neutral atoms (red dots, blue circles) trapped in a static \od optical lattice (gray boxes) can be rapidly rearranged via a two-step shuffling operation $\mathcal{R}$ (i)-(iii) facilitated by an auxiliary \od tweezer array (bottom left). Iterated shuffling and nearest-neighbor Rydberg interactions yield effective interactions on highly nonlocal coupling graphs such as the $m$-regular hypercube graph $Q_m$ (b). More generally, circuits (c) composed of shuffles (blue), nearest-neighbor controlled-$Z$ operations (red), and global rotations (purple) can be harnessed to generate Page-scrambled quantum states in $m$ iterations or strongly scrambling quantum channels in $2m$ iterations.
    }
    \label{fig:fig1}
\end{figure}

In this Letter, we propose experimental tools for achieving fast scrambling in near-term experiments with one-dimensional (1D) arrays of optically trapped neutral atoms \cite{weiss2004another,yang2016coherence,kim2016situ,gross2017quantum,bernien2017probing,zhang2017observation,levine2018high,levine2019parallel,kim2020quantum,scholl2020,ebadi2020quantum,young2020half}.
By rapidly shuffling atoms using optical tweezers it is possible to implement a broad family of nonlocal, sparsely coupled quantum circuits that realize fast scrambling quantum channels \cite{sekino2008,lashkari2011towards,hosur2016chaos}. 
%
Using such shuffling techniques for long-lived ground-state atomic qubits allows for the generation of highly nonlocal interaction graphs with only global rotations and nearest-neighbor Rydberg interactions. Though shuffling operations will generally be slower than Rydberg gates, limiting the number of gates to $\mathcal{O}[\log(N)]$ minimizes the primary sources of noise and decoherence, which arise from laser excitations to Rydberg levels \cite{levine2019parallel,lukin2001dipole,Heidemann2007,Jaksch2000,Mueller2014,saffman2010quantum}.
The simplest versions of these circuits efficiently produce many-body-entangled 
``graph states'' \cite{hein2006entanglement}, known computational resources for measurement-based quantum computation \cite{raussendorf2001aoneway,raussendorf2003measurement}, quantum metrology \cite{shettell2020graph}, quantum error correction \cite{looi2008quantum}, and quantum cryptography \cite{chen2004multi}. More sophisticated circuits built with the same experimental tools yield strongly scrambling quantum channels capable of robustly protecting quantum information against multiqubit erasure \cite{hayden2007black,yoshida2017efficient,Yoshida2019Disentangling,Bao2020a}.

Below we analyze iterated (Floquet) circuits built with these tools both for the idealized unitary case and the dissipative case expected in realistic implementations. We demonstrate that initially separable states can be Page scrambled using only $m \equiv \lceil \log_2 N \rceil$ nearest-neighbor interaction layers and construct deterministic circuits with only $2m$ interaction layers that strongly scramble quantum information regardless of the input state.



The basis for our protocol is the possibility to realize a family of sparse nonlocal coupling graphs via a quasi-\od shuffling procedure [Fig.~\ref{fig:fig1}(a)] on atoms in optical lattices facilitated by an auxiliary programmable \od tweezer array.
Straightforward stretching and interleaving tweezer operations \cite{beugnon2007two,endres2016atom,barredo2016atom,barredo2018synthetic} [Fig.~\ref{fig:fig1}(a),(i)-(iii)] can be used to rapidly shuffle the atomic positions. For $N = 8$ these motions execute the permutation
\begin{equation}
    \mathcal{R} = \begin{pmatrix}
    0 & 1 & 2 & 3 & 4 & 5 & 6 & 7 \\
    0 & 4 & 1 & 5 & 2 & 6 & 3 & 7
\end{pmatrix}
\end{equation}
with atoms labeled by $i = 0,1,\ldots,N-1$.
More generally, for system sizes $N = 2^m$ with $m$ an integer, a ``perfect'' shuffle or ``Faro shuffle'' operation \cite{diaconis1983mathematics,aldous1986shuffling} executes the nonlocal mapping
\begin{equation}
    i' = \mathcal{R}(i = b_m \ldots b_2 b_1) = b_1 b_m \ldots b_2,
\end{equation}
which cyclically permutes the bit order of the atomic index $i = b_m \ldots b_2 b_1$ written in binary such that the least significant bit $b_1$ of $i$ becomes the most significant bit of $\mathcal{R}(i)$. The shuffling operation $\mathcal{R}$, along with its inverse $\mathcal{R}^{-1}$ and generalizations thereof (see Supplemental Material \cite{SM}), are built on established tweezer-assisted techniques for defect removal in atom arrays \cite{endres2016atom,barredo2016atom,barredo2018synthetic} and can be implemented rapidly using a pair of acousto-optic deflectors (AOD) in crossed configuration and driven by independent rf signals $f_{x},f_{z}$ [Fig.~\ref{fig:fig1}(a), bottom left].


Repeated shuffling operations $\mathcal{R}$ dramatically rearrange the atomic positions. As a result, the propagation of quantum information is no longer constrained by the underlying 1D geometry of the fixed optical lattice. The simplest iterated circuit $\mathcal{E}_{Q_m} \equiv [\mathcal{R} \times \cze]^m$ generates effective controlled-$Z$ interactions on the $m$-regular hypercube graph $Q_m$ \cite{west2001introduction,bollobas2013modern}, a highly nonlocal, sparsely connected coupling graph  shown in Fig.~\ref{fig:fig1}(b). These nonlocal couplings allow many-body entanglement to be built up rapidly and efficiently using far fewer Rydberg interaction layers than would be needed in strictly 1D systems without shuffling. For example, given $N = 2^m$ qubits initialized in the product state $\prod_i \ket{+}_i = \prod_i (\ket{0} + \ket{1})_i / \sqrt{2}$ the circuit $\mathcal{E}_{Q_m}$ produces the Page-scrambled graph state $\ket{Q_m}$ after only $m$ interaction layers $\cze$ \cite{SM,shettell2020graph}.

\begin{figure}[t]
    \includegraphics[width=\columnwidth]{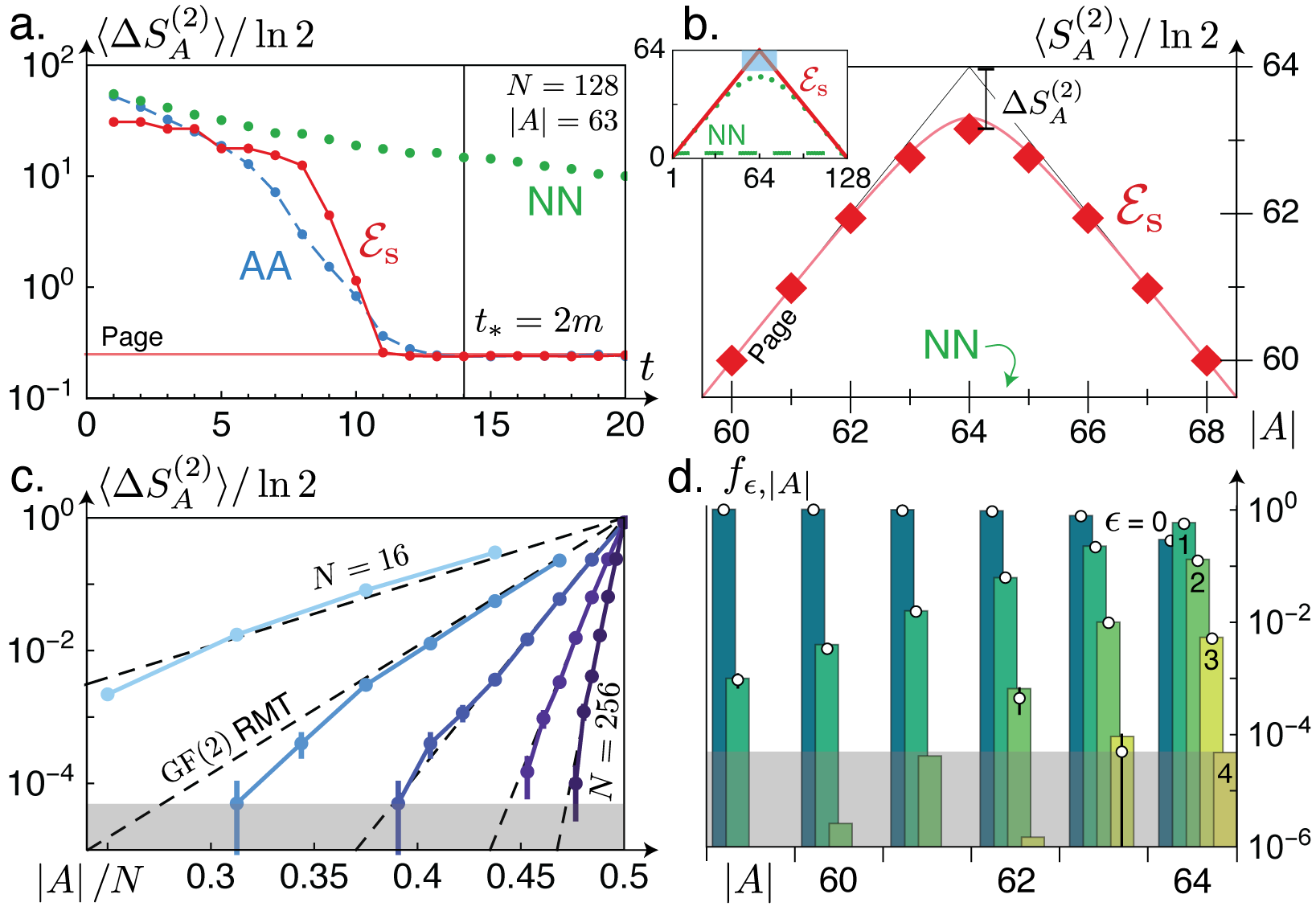}
    \caption{
        Page scrambling in $2m = 2\log_2 N$ steps.
        (a) The mean deficit $\langle \Delta S_A^{(2)} \rangle$ from volume-law entanglement entropy, sampled over $2 \times 10^4$ random bipartitions $A \cup \overline{A}$ of fixed size $\magn{A}$, decreases in the circuit $\chr$ on $N = 128$ qubits (solid red) at a rate comparable to a random all-to-all (AA) circuit (dashed blue) and much faster than a comparable nearest-neighbor (NN) circuit (dotted green). (b) After $2m$ circuit layers, the mean Renyi entropy $\langle \se_A \rangle$ (red diamonds), nearly saturates the Page curve (red), compared to a nearest-neighbor circuit of the same depth [(b), inset]. (c) The mean entropy deficit $\langle \Delta \se_A \rangle$ agrees with random matrix theory (dotted black) to within sampling fluctuations for  $N = 16,32,64,128,256$ (light to dark). (d) The fraction $f_{\epsilon,\magn{A}}$ of subsystems $A$ having less than maximal entanglement entropy (white dots) vanishes exponentially as a function of $\Delta \se_A / \ln 2 = \epsilon = 0,1,2,3$, in agreement with random matrix theory (vertical bars, light to dark). Error bars are shown or are smaller than markers; lines are guides to the eye; gray windows show statistical noise floor.
    }
    \label{fig:fig2}
\end{figure}

More sophisticated circuits built using the same experimental tools [Fig.~\ref{fig:fig1}(c)] can robustly scramble quantum information irrespective of the input state. By including global Hadamard $H$ and phase $P$ rotations, one can implement a strongly scrambling circuit
\begin{equation}
    \chr \equiv [\mathcal{R}^{-1} \czo H P]^m [\mathcal{R}^{-1} \cze H P]^m
\end{equation}
that yields widespread many-body entanglement after only $2m$ interaction layers $\cze,\czo$ for arbitrary input states, as demonstrated by numerical studies of Clifford circuits (Fig.~\ref{fig:fig2})\cite{SM,gottesman1998heisenberg,aaronson2004improved}. For $N = 128$ initially $z$-polarized qubits, randomly chosen subsystems $A$ consisting of an extensive number $\magn{A} = N/2-1$ of output qubits exhibit nearly maximal entanglement entropy after only $t_* = 2m = 14$ interaction layers, as measured by the Renyi entropy $S_A^{(2)} \equiv -\ln \tr{\rho_A^2}$ of the reduced density matrix $\rho_A \equiv \trover{\overline{A}}{\rho}$ [Fig.~\ref{fig:fig2}(a)]. The average deficit $\langle \Delta \se_A \rangle \equiv \magn{A} \ln 2 - \langle \se_A \rangle$ from perfect volume-law entanglement, sampled over $2 \times 10^4$ randomly chosen bipartitions $A \cup \overline{A}$ (solid red), rapidly decreases as a function of interaction layer $t$, saturating the Page limit $\Delta S_A^{(2)} = 2^{2 \magn{A} - N - 1}$ (horizontal red) \cite{Page1993,bianchi2019typical} prior to layer $t_* = 2m$. The timescale $t_* \sim \log N$ required for complete scrambling is comparable to that of a random all-to-all circuit (dashed blue) -- generally regarded as a prototypical fast scrambler \cite{sekino2008,lashkari2011towards,bentsen2019fast,gullans2020dynamical,Belyansky2020} -- and much shorter than for a nearest-neighbor circuit constructed without shuffling operations (dotted green).


In fact, the $2m$ interaction layers of the circuit $\chr$ suffice to generate volume-law mean entanglement entropy $\langle\se_A\rangle \approx \magn{A} \ln 2$ at all length scales $\magn{A} < N/2$ of the output state $\rho = \chr[\rho_0]$. Randomly chosen bipartitions $A \cup \overline{A}$, when organized by subsystem size $\magn{A}$, reveal a nearly ideal Page curve \cite{Page1993,bianchi2019typical} [Fig.~\ref{fig:fig2}(b), red].
The mean entanglement deficit $\langle \Delta \se_A \rangle$ is extremely small for almost all subsystem sizes and becomes substantial only for very large $\magn{A} \sim N/2$. Moreover, it is in excellent agreement with 
the predictions of random matrix theory (RMT) for binary matrices representing random stabilizer states over a range of system sizes [Fig.~\ref{fig:fig2}(c)] (see Supplemental Material \cite{SM}).

The widespread delocalization of information generated by the scrambling circuit $\chr$ is especially apparent when one considers how unlikely it is to find a subsystem $A$ of the output state $\rho$ with anything less than maximal entanglement [Fig.~\ref{fig:fig2}(d)]. Because the scrambling circuit $\chr$ consists entirely of gates chosen from the Clifford group, the Renyi entropy differs from its maximum value only by discrete bits $\Delta \se_A / \ln 2 = \epsilon = 0,1,2,\ldots$ \cite{gottesman1998heisenberg,aaronson2004improved}.
We therefore count the fraction $f_{\epsilon,\magn{A}}$ of the sampled bipartitions whose Renyi entropies differ from maximal by an amount $\epsilon$ [Fig.~\ref{fig:fig2}(d)]. We find that exponentially many subsystems $A$ have maximal entanglement entropy $\epsilon = 0$ (for $\magn{A} < N/2$), whereas it is exponentially rare to find a subsystem $A$ with entropy deficit $\epsilon > 0$.



\begin{figure}[b]
    \centering

    \includegraphics[width=\columnwidth]{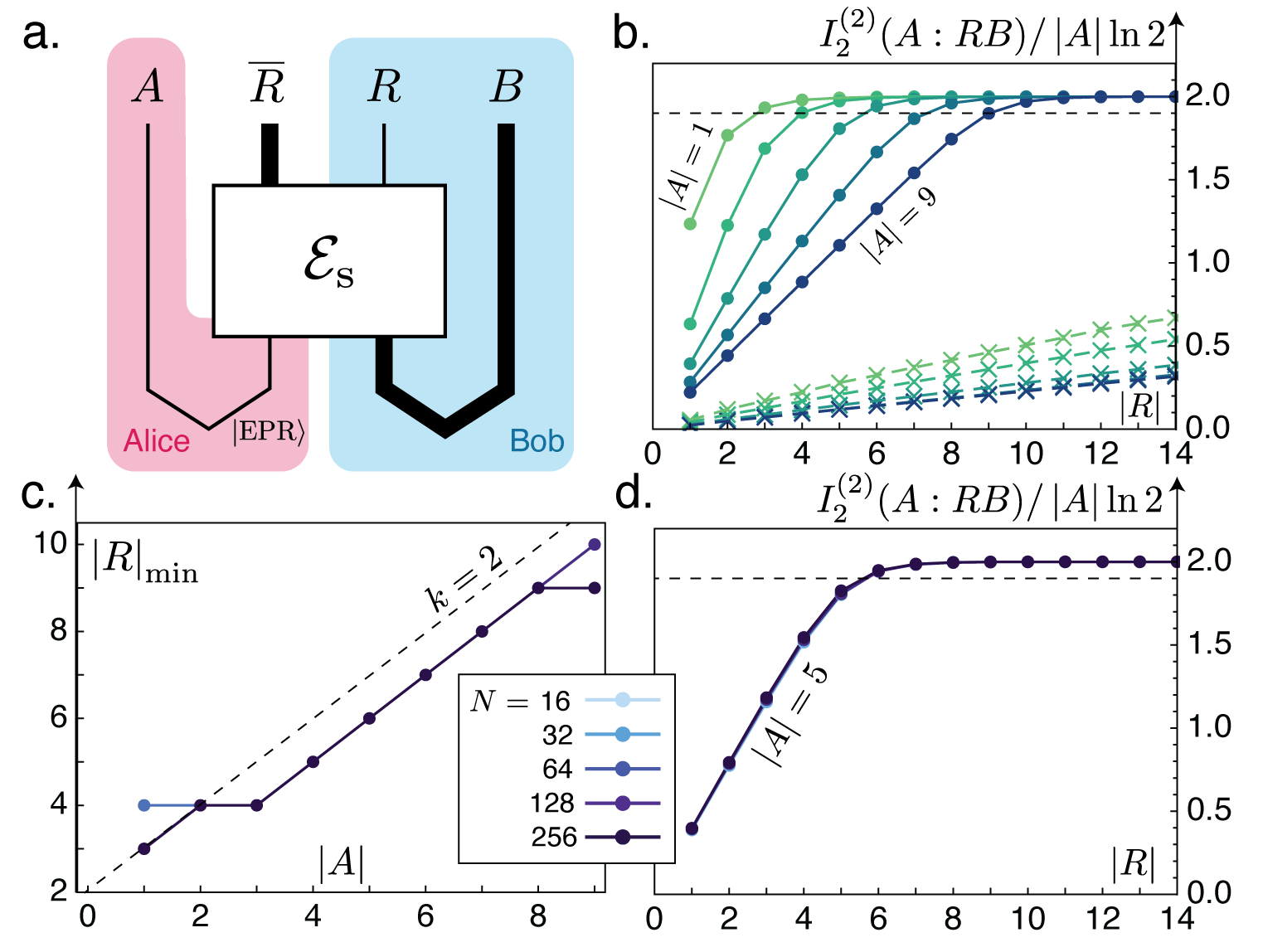}
    \caption{ Deterministic scrambling in the Hayden-Preskill thought experiment.
    (a) Scrambling in the circuit $\chr$ can be characterized by the mutual information $I^{(2)}_2(A:RB)$ between Alice's register $A$ (red) and Bob's registers $R,B$ (blue). (b) For $N = 128$ qubits, the mutual information grows rapidly as a function of Bob's output register $R$ over a range of message sizes $\magn{A} = 1,3,5,7,9$ (light to dark), saturating to within 5\% of its maximum value (dotted black) after Bob has collected only a handful $\magn{R}_{\mathrm{min}} \geq \magn{A} + k$ of output qubits with $k \leq 2$ [(c), dotted black]. Nearest-neighbor circuits of the same depth (crosses, dashed lines) show relatively low mutual information by comparison. (d) At fixed $\magn{A} = 5$, $I^{(2)}_2(A:RB)$ shows strong data collapse as a function of system size $N = 16,32,64,128,256$ (light to dark). Error bars smaller than markers; lines are guides to the eye.
    }

    \label{fig:fig3}
\end{figure}

Because of its ability to rapidly delocalize -- and thereby conceal -- quantum information, the scrambling circuit $\chr$ naturally serves a practical function in the context of quantum error correction and quantum communication.
In particular, strongly scrambling quantum channels are known to be excellent encoders that optimally protect quantum information against the effects of single-qubit erasure and other forms of local dissipation \cite{hayden2007black,yoshida2017efficient,gullans2020dynamical}. While prototypical examples of such encoding circuits are usually random, we demonstrate here that our deterministic circuit $\chr$ can be leveraged for precisely the same task, as illustrated by
the thought experiment of Hayden and Preskill \cite{hayden2007black,yoshida2017efficient,Yoshida2019Disentangling,Bao2020a} (Fig.~\ref{fig:fig3}).
Here, quantum information held by a local observer Alice $A$ is maximally entangled with the strongly scrambling quantum channel $\mathcal{E}_s$ using a collection of Einstein-Podolsky-Rosen (EPR) pairs. This information is subsequently recovered
with high fidelity by a maximally entangled observer Bob after measuring only a small subset $R$ of the output qubits and neglecting the rest $\overline{R}$. High fidelity teleportation of Alice's quantum information to Bob's register $B$ occurs if and only if the unitary channel is strongly scrambling \cite{hayden2007black,Yoshida2019Disentangling} and therefore presents a sharp criterion for diagnosing the presence of scrambling dynamics in our circuit $\chr$.


From the perspective of quantum error correction, we view the scrambling circuit $\chr$ as an encoding circuit that optimally protects Alice's information against erasure, allowing Bob to successfully reconstruct Alice's state even after discarding the large majority of output qubits $\overline{R}$. This is guaranteed, in principle, by large bipartite mutual information
\begin{equation}
    I^{(2)}_2(A:RB) = \se_A + \se_{RB} - \se_{ARB}
\end{equation}
between the qubits $A$ in Alice's control and those $R,B$ in Bob's control [Fig.~\ref{fig:fig3}(a)]. Numerical calculations with Clifford circuits demonstrate that the circuit $\chr$ on $N = 128$ qubits performs quite well as an encoding channel: the mutual information increases linearly with the number of output qubits $\magn{R}$ collected by Bob [Fig.~\ref{fig:fig3}(b)] and rapidly saturates to within 5\% of its maximum value $I^{(2)}_2(A:RB) = 2 \magn{A} \ln 2$ after he has collected a few more than $\magn{A}$ qubits. Physically, this implies that Bob need only gather a few $\magn{R}_{\text{min}} \geq \magn{A} + k$ of the output qubits in order to successfully decode Alice's message [Fig.~\ref{fig:fig3}(c)], with $k \leq 2$ for large $N$. By contrast, nearest-neighbor circuits of the same depth [Fig.~\ref{fig:fig3}(b), dashed lines] show low mutual information over a large range of output qubits $\magn{R}$. For fixed message size $\magn{A} = 5$, the mutual information shows strong data collapse as a function of system size $N$ [Fig. \ref{fig:fig3}(d)], indicating robustness to finite-size effects.

The numerical evidence presented in Figs. \ref{fig:fig2} and \ref{fig:fig3} demonstrates that $\chr$ is a fast scrambler in the ideal unitary case. Any realistic implementation of this scrambling circuit, however, must contend with the effects of noise and dissipation that will inevitably degrade its performance. In the following, we analyze a possible experimental realization of $\chr$ in detail, including the effects of decoherence to characterize its scrambling properties in a realistic setup.

We propose to use long-lived ground states $\ket{0},\ket{1}$ of neutral atoms as qubit states \cite{saffman2010quantum, Henriet2020, morgado2020quantum}. Single-qubit rotations allow for implementation of Hadamard and phase gates. By exciting $\ket{1}$ to a Rydberg state, controlled-$Z$ gates between neighboring atoms can be realized using strong van der Waals interactions \cite{lukin2001dipole, Heidemann2007, theis2016high, Jaksch2000, isenhower2010demonstration, Mueller2014, levine2019parallel, Madjarov2020}. Current experiments already achieve Rydberg gate fidelities~$> 0.99$ \cite{levine2019parallel, Madjarov2020}. A primary advantage of these operations is that they may be applied in parallel, using global optical or rf pulses. For our simulations, we take into account cross talk between atoms separated by the distance $r$, resulting from the $1/r^6$ decay of the van der Waals interaction. We model decoherence as dephasing noise with error rate $p$ per atom after each interaction layer (see Supplemental Material \cite{SM}).

\begin{figure}[t]
    \centering
    \includegraphics[width=\columnwidth]{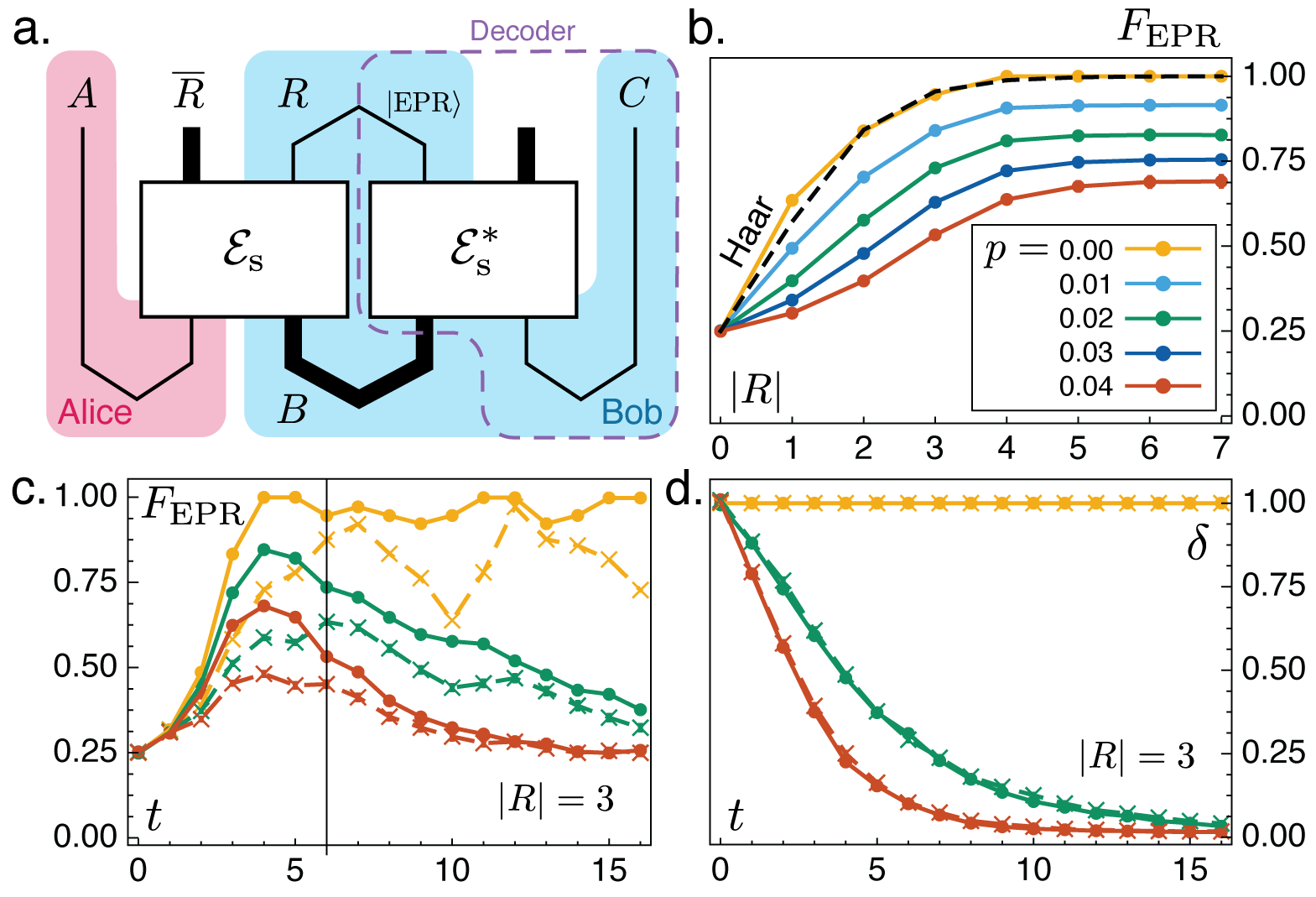}
    \caption{Information scrambling in the presence of dissipation. (a) Scrambling in the circuit $\chr$ is diagnosed by the fidelity $F_{\mathrm{EPR}}$ of recovering Alice's quantum information $A$ on Bob's register $C$ using a probabilistic decoding circuit (dotted purple). (b) For $N = 8$ at fixed circuit depth $t = 6$, the fidelity grows with the number of qubits $\magn{R}$ used in the decoder, indicating successful teleportation of Alice's information with fidelity $> 50 \%$ even in the presence of single-qubit errors at rates $p = 0.00,0.01,\ldots,0.04$ per 2-qubit gate (light to dark). For $p = 0$, the fidelity is nearly identical to that of a Haar-random circuit (dotted black). (c) The fidelity (dots, solid lines) grows with circuit depth $t$ and substantially outperforms nearest-neighbor circuits of the same depth (crosses, dotted lines) in the presence of dissipation. (d) The dissipation parameter $\delta$ falls as a function of circuit depth in both the scrambling circuit and nearest-neighbor circuit.
    Each data point averaged over $6 \times 10^4$ quantum trajectories, with error bars smaller than markers; lines are guides to the eye.}
    \label{fig:fig4}
\end{figure}

To distinguish between scrambling and decoherence, we attempt to recover Alice's information using a probabilistic decoding circuit [Fig.~\ref{fig:fig4}(a), dotted purple], following the scheme of Yoshida \emph{et al.} \cite{yoshida2017efficient,Yoshida2019Disentangling,Bao2020a}. This decoder consists of a complex-conjugated copy of the scrambling circuit and the ability to measure EPR pairs; decoding protocols of this type have been realized in pioneering experiments with trapped ions \cite{landsman2019verified}.
In the unitary case $p = 0$, the circuit decodes Alice's quantum information with a fidelity $F_{\EPR}=2^{I^{(2)}_2(A:RB) - 2 \magn{A}}$, conditioned on successful detection of $\magn{R}$ EPR pairs by Bob with probability $P_{\EPR}=2^{-I^{(2)}_2(A:RB)}$ [Fig.~\ref{fig:fig4}(b)]. 
Bob's ability to recover Alice's information is degraded by decoherence $p > 0$, where the product $\delta \equiv P_{\EPR} F_{\EPR} 2^{2 \magn{A}} \leq 1$ gives a natural metric for the strength of decoherence \cite{Yoshida2019Disentangling,Bao2020a}.

We compare the scrambling circuit  to an analogous circuit without shuffling and thus with controlled-$Z$ gates between nearest neighbors only. Notably, the nearest-neighbor circuit requires a longer time, measured in the number of interaction layers, to accomplish scrambling. While the decoherence metric $\delta$ behaves the same for the nearest-neighbor circuit and the scrambling circuit $\chr$ [Fig.~\ref{fig:fig4}(d)], for $p > 0$, the reachable teleportation fidelity $F_{\EPR}$ is significantly smaller for the slow scrambling nearest-neighbor circuit [Fig.~\ref{fig:fig4}(c)]. This demonstrates that fast scrambling is crucial in non-error-corrected systems, precisely because fewer gates provide fewer opportunities for dissipation. Our scrambling circuit $\chr$ is optimal in this regard as it generates strong scrambling using the minimal number of interaction layers $2m \sim \mathcal{O}(\log N)$ allowed by the fast scrambling conjecture
\cite{sekino2008,lashkari2011towards,bentsen2019fast}.

We have shown how deterministic, highly nonlocal iterated (Floquet) circuits can generate fast scrambling dynamics in a way that is amenable to direct experimental realization using fast shuffle operations on neutral atom qubits. This technique allows for rapid long-range spreading of entanglement, while minimizing errors from excitation of atoms to Rydberg states, and uses only shuffling operations, global single-qubit rotations, and parallel nearest-neighbor interactions. Building fast scrambling circuits in the laboratory opens connections to a wide range of ongoing areas, including fundamental limits on the spreading of quantum information \cite{hastings2006spectral,sekino2008,lashkari2011towards,bentsen2019fast}, experimental studies of toy models of black holes \cite{pikulin2017black,chew2017approximating,danshita2017creating,bentsen2019treelike,schuster2021many}, efficient encoders for quantum error-correcting codes \cite{hayden2007black}, and highly entangled resources for quantum computation \cite{raussendorf2001aoneway,raussendorf2003measurement}.
\nocite{iaconis2021quantum,Nielsen2009,Selinger2015,Nahum2017,kolchin1998,hardy1975AnIntroduction,Qiskit}
While we simulate example cases with stabilizer states \cite{gottesman1998heisenberg,aaronson2004improved,iaconis2021quantum,blake2020quantum} for large system sizes, analogous circuits built in the laboratory may employ arbitrary quantum rotations, exploring the complete many-body Hilbert space. We note that these graphs might also be constructed by other means, including collisional gate implementations for neutral atoms, or via direct wiring of hypercubic coupling graphs in superconducting qubit systems.

We recently became aware of a proposal \cite{schuster2021many} for further explorations of many-body quantum teleportation, based around nearest-neighbor Rydberg models with scrambling times $t_* \propto N$. The protocols we describe here for fast scrambling could be immediately combined with these interesting proposals to extend the example from Fig. \ref{fig:fig4} discussed here. The data for this manuscript is available in open access at \cite{hashizume2021data}.

\section{acknowledgments}

We thank Jon Pritchard, Monika Schleier-Smith, Hans Peter B\"{u}chler, and Simon Evered for stimulating and helpful discussions. GSB is supported by the DOE GeoFlow program (DE-SC0019380). Work at the University of Strathclyde was supported by the EPSRC Programme Grant DesOEQ (EP/P009565/1), the EPSRC Quantum Technologies Hub for Quantum Computing and simulation (EP/T001062/1), the European Union’s Horizon 2020 research and innovation program under grant agreement No. 817482 PASQuanS, and AFOSR grant number FA9550-18-1-0064. SW was supported by the European Union under the ERC
consolidator grant SIRPOL (grant N. 681208).

\bibliography{rydberg_formatted}

\clearpage
\onecolumngrid

\renewcommand{\thefigure}{S\arabic{figure}}
\renewcommand{\theequation}{S\arabic{equation}}

\setcounter{equation}{0}
\setcounter{figure}{0}
\section{Deterministic Fast Scrambling with Neutral Atom Arrays:\\Supplemental Material}

\section{I. Shuffling Operations with Optical Tweezers}

\begin{figure}[b]
    \centering
    \includegraphics[width=0.45\textwidth]{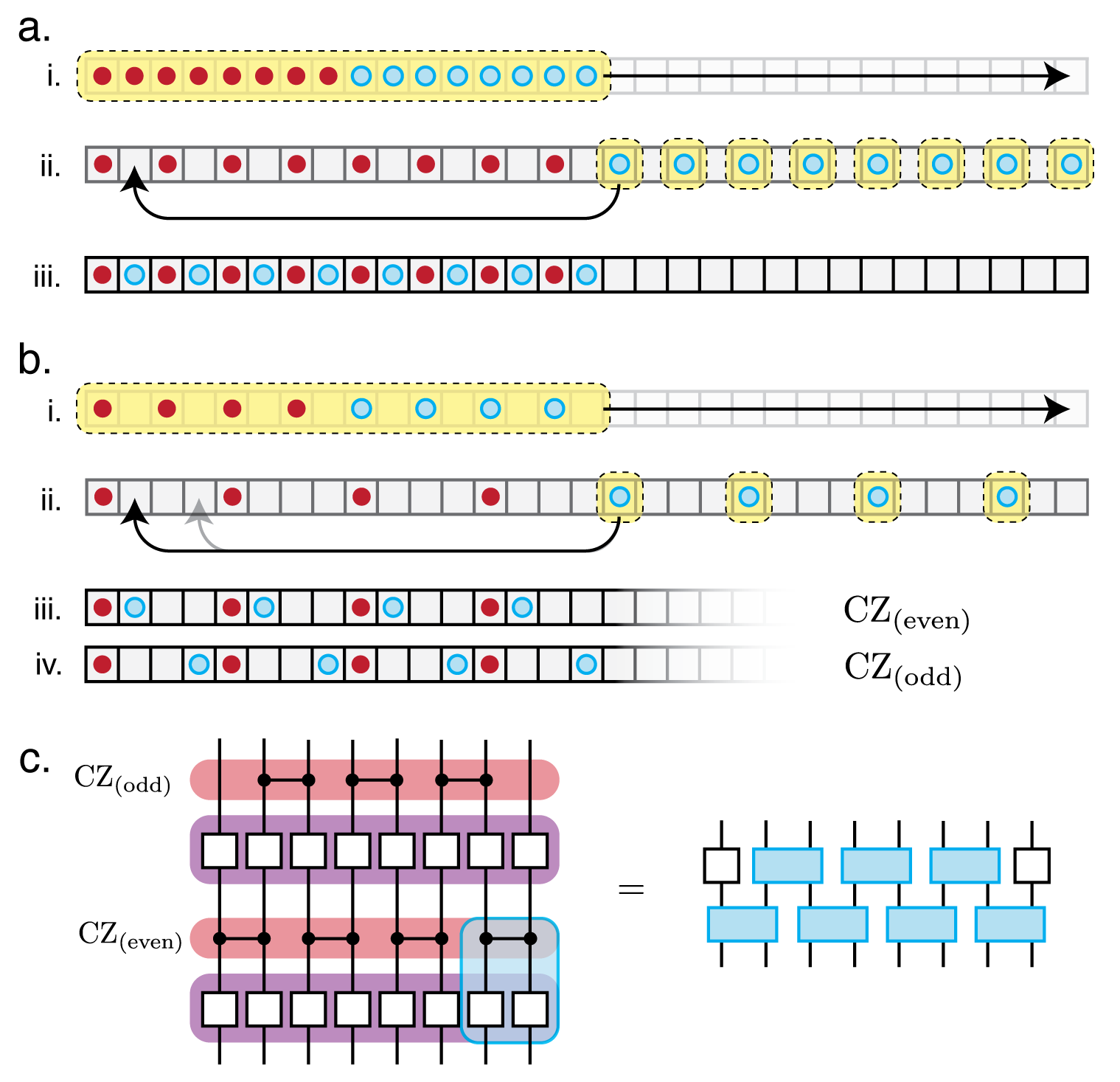}
    \caption{\textbf{Quasi-\od Shuffling Operations} (a) Uniform-density shuffling procedure $\mathcal{R}$ for $N = 16$ atoms, similar to main text. (b) Even- and odd-shifted shuffling procedures for $N = 8$ allow subsequent controlled-$Z$ gates to be applied preferentially on even (iii) or odd (iv) bonds of the \od lattice, respectively. (c) When combined with global single-qubit rotations (purple), even- and odd-bond gates (red) can be used to generate bricklayer circuits composed of non-commuting two-qubit gates (blue).}
    \label{fig:tweezerrearrange}
\end{figure}

A \emph{Faro shuffle} or \emph{perfect shuffle} $\mathcal{R}$ (Fig.~\ref{fig:tweezerrearrange}a) begins with $N$ atoms (red dots, blue circles) labeled $i = 0,1,\ldots,N-1$ initially trapped at sites $x(i) = i a$ of a fixed \od optical lattice (gray boxes) with spacing $a$. An additional $N$ empty sites $x = N a, \ldots, (2N-1) a$ are reserved for `scratch space.'
An auxiliary \od tweezer array superimposed on the fixed lattice captures all $N$ atoms and performs an adiabatic row-stretch operation that relocates atoms at site $x$ to site $2 x$ (Fig.~\ref{fig:tweezerrearrange}a(i)). The first $N/2$ tweezers are then switched off, allowing the atoms on the left half of the cloud (Fig.~\ref{fig:tweezerrearrange}a(ii), red dots) to be recaptured by the fixed \od lattice. The remaining $N/2$ tweezers adiabatically transport the atoms in the right half of the cloud (Fig.~\ref{fig:tweezerrearrange}a(ii), blue circles) below the trap array, leftward by a distance $\Delta x = -(N-1)a$, and back into the trap. Upon switching off the tweezer array, the atoms $i$ are rearranged in the linear trap by the permutation $i' = \mathcal{R}(i)$ (Fig.~\ref{fig:tweezerrearrange}a(iii)). The inverse operation $\mathcal{R}^{-1}$ is executed by simply reversing the above steps.

The simplest shuffling operation $\mathcal{R}$ deposits the atoms into the optical lattice with uniform density, leading to translation-invariant nearest-neighbor Ising couplings in the subsequent interaction step. By slightly modifying the original scheme the atoms may be deposited into the optical lattice non-uniformly such that subsequent Rydberg-Rydberg interactions generate controlled-$Z$ gates only between nearest-neighbor pairs (Fig.~\ref{fig:tweezerrearrange}b), while atoms separated by more than one lattice site have negligible interactions. In particular, by using a rarified atomic array and preferentially shifting the transported atoms left or right by one lattice site at the end of the interleaving step (Fig.~\ref{fig:tweezerrearrange}b(ii)), controlled-$Z$ gates may be placed only on even (Fig.~\ref{fig:tweezerrearrange}b(iii)) or odd (Fig.~\ref{fig:tweezerrearrange}b(iv)) bonds of the expanded 1D lattice. These even- and odd- interaction layers, along with global Hadamard and Phase gates, can be used to readily construct `bricklayer' circuits of non-commuting gates as shown in Fig.~\ref{fig:tweezerrearrange}c.


The shuffling operations described here are quasi-1D in that they maintain the one-dimensional linear geometry of the right-hand atoms (blue circles) during transport and utilize only straightforward spatial translations and row-stretch operations of the auxiliary tweezer array. These operations are straightforward to implement by sweeping the frequencies of the tweezer sidebands in the independent RF drive signals $f_x,f_z$. The speed of these frequency sweeps and the accompanying spatial motion of the atoms will be limited by adiabaticity; recent experimental work has demonstrated adiabatic transport speeds up to $\si{75 \um/\ms}$, suggesting that single shuffling operations $\mathcal{R}$ can be realistically executed on timescales of order $\si{1 \ms}$ for $N = 32$ or more atoms in optical lattices with spacing $a = \si{3 \um}$ \cite{ebadi2020quantum}.

\section{II. Graph States}

Given an undirected graph $G = (V,E)$ with vertices $i,j \in V$ and edges $(i,j) \in E$, a \emph{graph state} is defined as
\begin{equation}
    \ket{G} = \prod_{(i,j) \in E} \mathrm{CZ}_{(i,j)} \ket{+}^V
\end{equation}
where $\ket{+}^V = \prod_{i \in V} (\ket{0} + \ket{1})/\sqrt{2}$ and $\mathrm{CZ}_{(i,j)}$ is a controlled-$Z$ gate applied between qubits $i,j$ that live on the vertices of the graph \cite{hein2006entanglement}. This state is generated by preparing $N$ qubits in $\ket{0}$, applying a global Hadamard rotation, and applying a controlled-$Z$ gate between sites $i,j$ for each edge $(i,j) \in E$. Because the $\mathrm{CZ}_{(i,j)}$ gates mutually commute, the edges in the graph have no preferred ordering and every undirected graph $G$ is in one-to-one correspondence with a graph state $\ket{G}$.

The entanglement properties of the graph state $\ket{G}$ can be immediately extracted from the adjacency matrix $\Gamma_{ij}$ of the graph $G$:
\begin{equation}
    \Gamma_{ij} = \begin{cases} 
      1 & (i,j) \in E\\
      0 & \mathrm{otherwise}
   \end{cases}
\end{equation}
which encodes the graph connectivity as a binary matrix. For an arbitrary bipartition $A \cup \overline{A}$, we may reorder the rows and columns of $\Gamma_{ij}$ to bring it into block form
\begin{equation}
    \Gamma_{ij} = \begin{bmatrix}
\Gamma_{AA} & \Gamma_{A\overline{A}}\\
\left(\Gamma_{A\overline{A}}\right)^T & \Gamma_{\overline{A}\overline{A}}
\end{bmatrix}
\end{equation}
where the off-diagonal sub-matrix $\Gamma_{A \overline{A}}$ represents the edges connecting regions $A,\overline{A}$. Then the Schmidt rank $r(\rho_A)$ of the reduced density matrix is given by the binary rank (i.e. rank over the field $\gft$) of the sub-adjacency matrix \cite{hein2006entanglement}
\begin{equation}
    r(\rho_A) = \mathrm{rank}_{\gft}(\Gamma_{A\overline{A}}) \label{gf2rank}
\end{equation}
From the Schmidt rank we can compute entropy measures such as the Renyi-2 entropy. In particular, for graph states and stabilizer states the Renyi-2 entropy is simply $S_A^{(2)} = -\ln \tr{\rho_A^2} = r(\rho_A) \ln 2$.

\begin{figure}[h]
    \centering
    \includegraphics[width=0.6\textwidth]{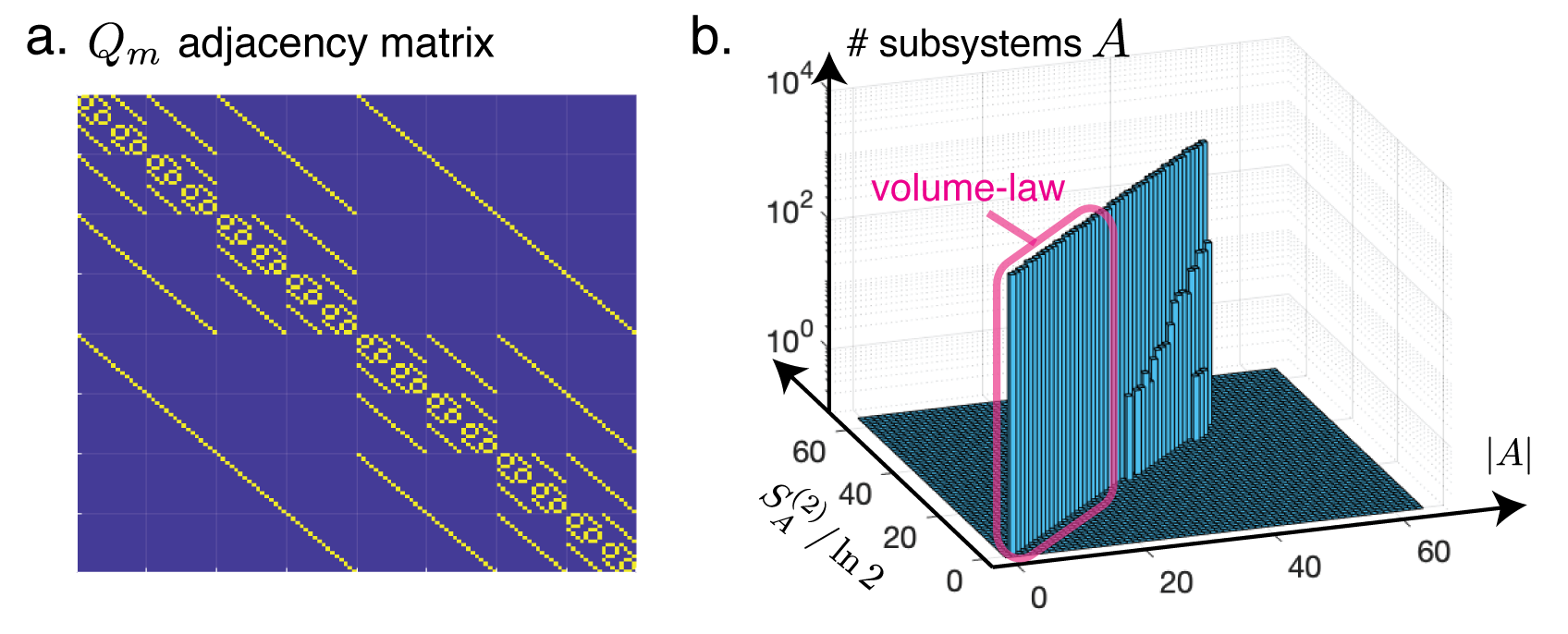}
    \caption{\textbf{Hypercube graph state $\ket{Q_m}$.} (a) Adjacency matrix $\Gamma_{ij}$ for the hypercube graph $Q_m$ on $N = 128$ vertices. (b) $10^5$ randomly-chosen subsystems $A$, arranged by size $\magn{A}$, exhibit perfect volume-law entanglement (pink box) for all subsystems up to $\magn{A} \approx N/8$. For larger subsystems $\magn{A} \gtrsim N/8$ we find some subsystems $A$ with entropy deficit $\epsilon > 1,2$, but the average entropy is still dominated by volume-law behavior.}
    \label{fig:qmmatrix}
\end{figure}

Graph states with sufficient nonlocal connectivity in the graph $G$ may be \emph{Page entangled}, meaning that they exhibit volume-law entanglement $S_A^{(2)} \propto \magn{A}$ in all bipartitions $A \cup \overline{A}$ of size $\magn{A} < \alpha N$ for some $\mathcal{O}(1)$ constant $\alpha$ \cite{sekino2008,lashkari2011towards}.
As discussed above, volume-law entanglement in the reduced state $\rho_A$ is equivalent to the off-diagonal adjacency matrix $\Gamma_{A \overline{A}}$ having full rank. The graph state $\ket{Q_m}$, whose adjacency matrix is plotted in Fig.~\ref{fig:qmmatrix}a, is Page-scrambled with $\alpha \gtrsim 1/8$ as demonstrated by computing the rank over random samples of the submatrix $\Gamma_{A\overline{A}}$ (Fig~\ref{fig:qmmatrix}b).




\section{III. Stabilizer States\label{sup:Cliford}}
The idea of stabilizer formalism is to specify a state, $\textit{stabilizer state}$,
as a simultaneous eigenstate of a set of operators called \textit{stabilizers}.
A state $\ket{\Sigma}$ is said to be stabilized by a stabilizer $\mathcal{O}$ 
if $\ket{\Sigma}$ is a $+1$ eigenstate of $\mathcal{O}$ as an action of $\mathcal{O}$ on $\ket{\Sigma}$ does not change the state. 
Such a set of operator can be constructed by making use of the properties of Pauli group on $N$ qubits, $\mathcal{P}^{N}$.
$\mathcal{P}^{N}$ is a group consists of all the possible tensor products of 
the Pauli operators ($I$,$X$,$Y$,$Z$) and phases ($1$,$i$,$-1$,$-i$) 
For example, $-I \otimes X \otimes Z \otimes I \otimes Y$ is an element, also referred to as a \textit{Pauli string}, 
of a group $\mathcal{P}^{5}$.
An operator, $\mathcal{O}_{\ket{\Sigma}}$, which stabilizes and uniquely defines $\ket{\Sigma}$ 
is constructed from a set of $N$ linearly independent Pauli strings, $\mathcal{P}^{N}_{\Sigma} \in \mathcal{P}^{N}$, 
that stabilize the state $\ket{\Sigma}$, as follows
\begin{align}
    \mathcal{O}_{\Sigma}=\frac{1}{2^N}\prod_{P \in \mathcal{P}^{N}_{\Sigma} }^{N} \left(P + I\right), \label{stabilizerO}
\end{align}
where the linear independence of Pauli string means a Pauli string $P_i \in P^{N}_{\Sigma}$ 
cannot be represented as a product of $ P_{j} $ ($j\neq i$) 
and $\ket{\Sigma}$ is a unique simultaneous eigenstate of the strings 
of $\mathcal{P}^{N}_{\Sigma}$\cite{gottesman1998heisenberg,aaronson2004improved,Nielsen2009,Selinger2015}.

\subsection{A. Classical Simulation of Stabilizer States in Clifford Circuits}
Consider a unitary operator $\mathcal{O}_C$ which transforms an element of Pauli group to some element in the same group.
The action of such unitary on a stabilizer state $\ket{\Sigma}$ produces a stabilizer state $\ket{\Sigma'}$,
which is stabilized by an operator $\mathcal{O}_{\Sigma'} = \mathcal{O}_C \mathcal{O}_{\Sigma} \mathcal{O}_C^{\dagger}$.
A set of operators $\{ \mathcal{O}_C \} \equiv \mathcal{C}^{N} $ with such properties forms Clifford group on $N$ qubits.
The evolution of a stabilizer state under actions of the elements of the Clifford group can therefore be simulated 
by keeping track of how $N$ Pauli strings in the initial stabilizer are being mapped.

A stabilizer state undergoing unitary evolution by the elements in the 
Clifford group known to be able to compute classically in polynomial time 
as proven by Gottesman and Knill\cite{gottesman1998heisenberg,aaronson2004improved}.
The simulation of the stabilizer states in Clifford circuit can be done efficiently with a logical operation on a 
$N$ by $2N+1$ binary matrix $M$\cite{aaronson2004improved}. 
In this binary matrix, each row represents a Pauli String. 
For each row, $l$\textsuperscript{th} character of the string is encoded by mapping the number 
of $X$ to $l$\textsuperscript{th} column and the number of $Z$ to $N+l$\textsuperscript{th} column.
Consequently, having $Y=iXZ$ corresponds to $1,1$ on $l$\textsuperscript{th} and $N+l$\textsuperscript{th} columns.
The last column is reserved for tracking the overall phase of the element, where $0$ corresponds to $-1$, and $1$ corresponds to $+1$. 
As Clifford gate transforms one Pauli string to another, evolution of a stabilizer state $\ket{\Sigma}$ 
can be kept track by altering the strings of initial $P^{N}_{\Sigma}$ accordingly 
to the transformation rules\cite{aaronson2004improved}.

It is known that any $N$-qubit Clifford group can be generated from combinations 
of Hadamard ($H$), Phase ($P$) and Controlled-NOT (C-NOT) gates \cite{gottesman1998heisenberg,aaronson2004improved,Selinger2015}
acting on different sets of qubits in the system. 
Hadamard gate acting on a site maps operators $Z$ to $X$ and $X$ to $Z$ in a Pauli string.
Phase gate acting on a site, on the other hand, maps $X$ to $Y$ and $Y$ to $X$.
C-NOT gate, acting on two qubits, control qubit $l$ and target qubit $m$,
flips the target basis whenever the state of the $l$ is $\ket{1}$, i.e.
\begin{align}
    \text{C-NOT}_{l,m} = \frac{1}{2}\left( (I_l-Z_l) + (I_l+Z_l)X_m) \right).
\end{align}
By computing the unitary transformation on all the element in the 2-qubit Pauli group, one finds that
there are only 4 non-trivial transformation rules:
\begin{align}
    X_lI_m\ \text{to} \ X_lX_m, \ I_lX_m \ \text{to} \ I_lX_m, \ Z_lI_m\ \text{to} \ Z_lI_m, \ I_lZ_m \ \text{to} \ Z_lZ_m.
\end{align}

\subsection{B. Construction of Random Nearest Neighbor and Random All-to-All Models}
Two random circuits that are investigated in this paper in accordance to Fig.~\ref{fig:fig2}a 
are random nearest-neighbor circuit and random all-to-all curcuit.
The random nearest-neighbor circuit is simulated by acting randomly chosen gates from 
the $2$-qubit Clifford group (including single site rotations and identities) on even bonds at even interaction layers, 
and odd bonds at odd interaction layers, making a brickwork pattern. 
Random all-to-all circuit, on the other hand, for each interaction layer, the randomly chosen gates
are applied to the even bonds and after the parallel application of $N/2$ gates on those bonds, sites are permuted randomly.
This construction makes one time step -- an interaction layer -- to be defined unambiguously as a parallel application of 
$N/2$ gates on $N/2$ non-overlapping pairs of qubits in the system across the circuits discussed in the main text.
A complete set of elements of $2$-qubit Clifford group is generated from the elementary gates (H,P,C-NOT)
with the algorithm formulated by Selinger\cite{Selinger2015}.

\subsection{C. Entanglement Entropy of Stabilizer States}
In Sec.~III. A, 
the stabilizer state $\ket{\Sigma}$ is defined with a set of $N$ linearly independent Pauli strings $\{ P_i \}$ which stabilizes 
$\ket{\Sigma}$ and $\ket{\Sigma}$ being the unique simultaneous eigenstate of the strings.
The uniqueness of $\ket{\Sigma}$ and the property which $P_i$ having only two, namely $+1$ and $-1$, 
$\mathcal{O}_{\Sigma}=\frac{1}{2^N}\prod_{P \in \mathcal{P}^{N}_{\Sigma}} \left(P + I\right)$ (Eq.~\eqref{stabilizerO}) 
is a projector that projects onto a subspace $\ket{\Sigma}\bra{\Sigma}$, 
or equivalently it is the density matrix of the state $\ket{\Sigma}$.
Knowing the density matrix, entanglement entropy of any subsystem $A$ of $\ket{\Sigma}$ 
can be calculated readily\cite{Nahum2017,gullans2020dynamical}.
Expanding the products in $\mathcal{O}_{\Sigma}$,
\begin{align}
    \mathcal{O}_{\Sigma} = \frac{1}{2^{N}} \sum_{g\in\mathcal{G}} g,
\end{align}
where a set $\mathcal{G}$ is a set generated by all the possible products of $\{ P_i \}$ and an identity $I^{\otimes N}$,
$\mathcal{O}$ can be written as a sum of Pauli strings as Pauli Group is closed in multiplication.
The reduced density matrix on the subsystem $A$, $\rho_A$ is obtained by tracing out the complement of $A$, $\overline{A}$.
This is equivalent to throwing away the Pauli strings which the parts corresponding to $\overline{A}$ 
is not an identity as Pauli matrices are traceless except for the identity
\begin{align}
    \rho_{A} = \trover{\overline{A}}{\rho} 
    = \frac{2^{|\overline{A}|}}{2^{N}}\sum_{g_A\in\mathcal{G}_A} g_{A} = \frac{1}{2^{|A|}}\sum_{g_A\in\mathcal{G}_A} g_{A},
\end{align}
where $\mathcal{G}_A\in\mathcal{G}$ is a set of all $g$ with $\mathrm{Tr}_{\overline{A}}\{ g \} \neq 0$\cite{Nahum2017}.
Let $N_A$ be the number of linearly independent Pauli strings that generate $g_A$, then 
$\sum_{g_A\in\mathcal{G}_A}g_A$ is proportional to a projector of rank $2^{|A|-N_A}$ as this projector projects out the $-1$ 
eigenstates of its generators.
Therefore the von Neumann or Renyi entropy $S_A$ is 
\begin{align}
    S_A = \left(|A| - N_A \label{entropyN_A}\right)\ln 2
\end{align}
Using $2^{N_A+N_{\overline{A}}} = 2^{N}$ and equivalence of $N_{\overline{A}}$ to the binary rank of $M_{\overline{A}}$,
where $M$ is the binary matrix representing the state $\ket{\Sigma}$ 
and $M_{\overline{A}}$ is the matrix with columns of $M$ corresponding to the subsystem $\overline{A}$,
\begin{align}
    S_A = \left(\mathrm{rank}_{\gft}(M_{\overline{A}}) - |\overline{A}| \right)\ln 2.
\end{align}
From $S_A=S_{\overline{A}}$, we obtain
\begin{align}
    S_A = \left(\mathrm{rank}_{\gft}(M_A) - |A|\right)\ln 2.
\end{align}

\subsection{D. Average Entropy of a Subsystem of Random Stabilizer States}
As it is shown in the previous section, a stabilizer state of $N$ qubits can be represented by a binary matrix $M$ 
of the dimensions $N$ by $2N$.
A random stabilizer state can therefore be constructed from a random binary matrix with a constraint $\mathrm{rank}_{\gft}(M)=N$.
Also it is shown that the entropy of a subsystem of size $|A|$ of a stabilizer state can be obtained 
by subtracting $|A|$ from the binary rank of the binary matrix of the corresponding region.
The average entropy of a random subsystem $A$ in a random stabilizer state
can therefore be estimated by approximating the $\mathrm{rank}_{\gft}(M_A)$ with that of random binary matrices.

An $N$ by $2|A|$ ($2|A| < N$) random binary matrix $M_{|A|}$ can be constructed by appending $2|A|-1$ rows of random binary vectors 
to $N$ by $1$ matrix. Each time a new row is added, the rank does not increase with probability $2^{k}/2^{2|A|}$ where $k$ is 
the current rank, and the rank increases otherwise. 
The probability of the matrix to have exactly $r$ when the full matrix is constructed is, therefore,
\begin{align}
    P(\mathrm{rank}_{\gft}(M_{|A|})=r) = 
    \sum_{t\in \mathcal{T}} \prod_{i=1}^{r} \left(\frac{2^{i-1}}{2^{2|A|}} \right)^{t_{i}-t_{i-1}-1}
    \left(1-\frac{2^{i-1}}{2^{2|A|}}\right)
\end{align}
where $\mathcal{T}$ is a set of all the configurations of the row numbers where rank increases by $1$ and for all $t\in\mathcal{T}$,
$t_0=0$.
For large $|N|$, the above expression can be approximated by the following expression \cite{kolchin1998}
\begin{align}
    P\left( \mathrm{rank}_{\gft}(M_{|A|}) = 2|A|-\epsilon \right) &\approx 2^{-\epsilon\left( N-2|A|+\epsilon\right)}
    \times \prod_{i=\epsilon+1}^{\infty} \left( 1-\frac{1}{2^i} \right) 
    \prod_{i=1}^{N-2|A|+\epsilon}\left( 1-\frac{1}{2^{i}} \right)^{-1}.
\end{align}
Using this expression, the average entropy deficit of a subsystem of size $|A|$ 
of a random stabilizer state is approximated as follows
\begin{align}
    \langle  \Delta \se_{A} \rangle 
    = \left( \sum_{\epsilon} \epsilon P\left( \mathrm{rank}_{\gft}( M_{|A|} ) = 2|A|-\epsilon \right)\right)\ln 2
    \approx 2^{2|A| - N}\ln 2
\end{align}
where the approximation is made by only considering $\epsilon=0$ and $1$, and taking the limit of $1 \ll N-2|A|$.
This result coincides with but slightly larger than the expected entropy deficit 
of a Haar random state\cite{Page1993}, 
by a constant factor of $2\ln 2$, which is of the order $1$.
Also, a random binary matrix constructed in this way is almost guaranteed to have 
the rank of $N$ for $1\ll N$ as the probability of the binary rank of the matrix 
to be $N-\epsilon$ with $\epsilon=1$ is $2^{-N}$.
Thus the vast majority of random $N \times 2N$ binary matrices represent stabilizer states.


\section{IV. Average Entropy of a Subsystem of Area-Law States in 1-D Quantum Systems}

\begin{figure}
    \includegraphics[width=0.8\textwidth]{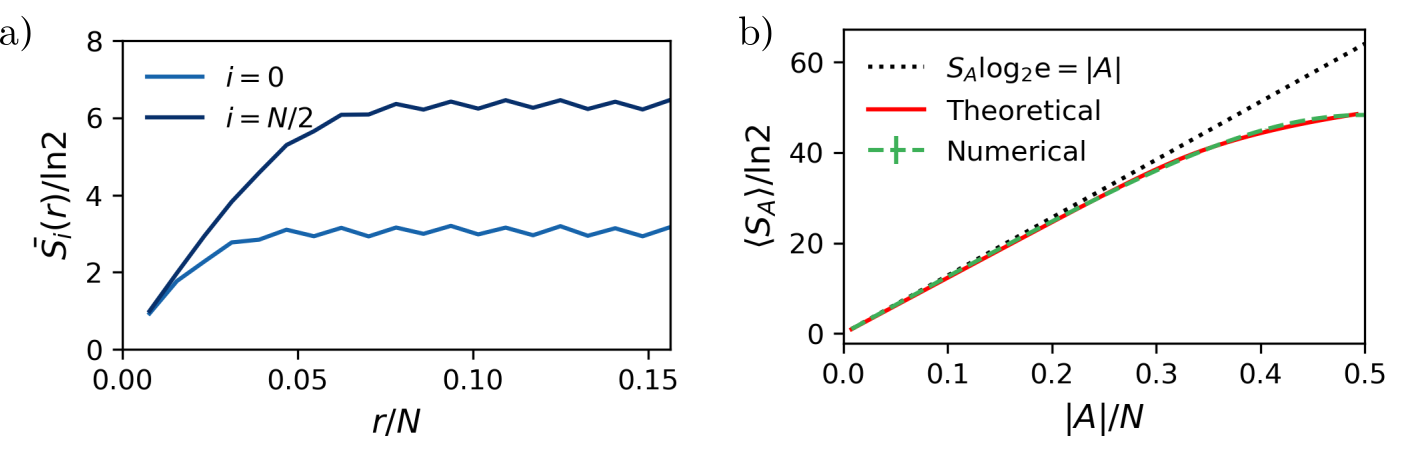}
    \caption{(a) The entanglement entropy of a region consists of $r$ consecutive qubits starting from site $i$, $S_i(r)$, 
        for $i=0$ and $i=N/2$ of the nearest-neighbor random Clifford circuit with open boundary condition for $N=128$ 
        at the number of interactions $t=2\log_2(N)=14$. 
        The bar on $\bar{S_i}(r)$ indicates that it is an averaged quantity over the different realizations of the random circuit.
        For this simulation the average of up to $1000$ realizations are taken. 
        (b) The average entropy of up to $2\mathrm{e}4$ random subsystems, $A$, 
        of an output state of a single trajectory of the random nearest-neighbor circuit ordered by the subsystem size $|A|$ (green dotted line). The theory line (red solid line) is computed using 
        Eq.~\eqref{SMNN:theory} with the entropy as function of $r$ consecutive regions approximated by 
        $S(r)\sim \tilde{S}(r) = \frac{1}{2}\left(\bar{S}_{0}(r) + \bar{S}_{N/2}(r) \right)$. Here the average of the functions 
        $\bar{S}_0(r)$ and $\bar{S}_{N/2}(r)$ are taken to take account of the effect from the open boundary condition.
        \label{SM_arealaw}}
\end{figure}

In this section, we derive the expression for average entropy of a subsystems $\mathcal{A}$ consisting of randomly chosen qubits 
with the subsystem size of $|A|$ drawn from the $N$-qubit state with area-law entanglement entropy.
Let the entanglement entropy of $r$ consecutive region from $i$\textsuperscript{th} qubit
to be expressed as $S_i(r)$. We assume the translational symmetry of this function such that
$S_i(r)=S_j(r)=S(r)$ also holds for sites $j \neq i$.
Given a particular subsystem and configuration $A \in \mathcal{A} $, one can always find $Q_A$ sets of $q_{A,k}$ ($k=1,2,\dots,Q_A$) 
qubits drawn from a consecutive region in the system. 
For example a set $A=\{1,5,6,7,10\}$ 
has $Q_A=3$ with $q_{A,1}=1$ (from $\{1\}$), $q_{A,2}=3$ (from $\{5,6,7\}$), and $q_{A,3}=1$ (from $\{10\}$).
Assuming that the mutual information between the two sets are $0$,
which is true for the vast majority of cases for the state with the area-law entanglement entropy for $|A| \ll N/2$,
one can write the entanglement entropy of a given configuration as $\sum_{k=1}^{Q_A} S(q_{A,k})$.
For fixed $|A|$ and $Q_A$, there are $C(Q_A)=\begin{pmatrix} N - |A| + 1 \\ Q_A\end{pmatrix}$ possible ways 
to draw $|A|$ qubits from $N$ qubits such that they have exactly $Q_A$ sets of $\{ q_{A,k} \}$ consecutive regions.
Finally there are $p(|A|)=Q_A$ ways to partition a subsystem $A$ into the cells which contains at least $1$ consecutive qubits,
where $p(|A|)$ is a well known function in number theory called \textit{partition function}
(not to be confused with the partition function from the thermodynamics) \cite{hardy1975AnIntroduction}.
From these, the expression for the average entropy of a random subsystem of size $|A|$ of an area-law entangled quantum state is given as
\begin{align}
    \langle S_{A} \rangle = 
    \frac{\sum_{l=1}^{p(|A|)} C(Q_{A_l}) \sum_{k=1}^{Q_{A_l}} S(q_{A_{l},k})}{\sum_{l=1}^{p(|A|)}\sum_{k=1}^{Q_{A_l}} C(Q_{A_l})} 
    \label{SMNN:theory}
\end{align}
where $A_l$ ($l=1,2,\dots,p(|A|)$) goes through all the different ways to partition the subsystems.
As shown in Fig.~\ref{SM_arealaw}b, the average entropy of the random bipartitions as a function of the subsystem size 
result of the numerical simulation of random nearest-neighbor circuit for $N=128$ at $t_*=2m=14$ 
shows an excellent agreement with the theoretical values computed explicitly with Eq.~\ref{SMNN:theory} 
with $S(r)=\tilde{S}(r)=\frac{1}{2}\left( \bar{S}_{0}(r)+\bar{S}_{N/2}(r) \right)$ where $\bar{S}_0(r)$ and $\bar{S}_{N/2}(r)$ 
are estimated by averaging up to $1\times10^{3}$ realizations of the circuit (Fig.~\ref{SM_arealaw}a).

\section{V. Characterizing Scrambling via the Hayden-Preskill Experiment} \label{sec:aydenpreskill}
    %

Originally conceived in the context of the black hole information problem, the Hayden-Preskill thought experiment (Fig.~\ref{fig:fig3}) can be viewed as a general conceptual tool useful for characterizing the scrambling properties of quantum channels $\mathcal{E}$. We focus for the moment on unitary channels $\mathcal{E} = U$ and consider a local observer Alice who wishes to use this channel to encode some quantum information. To do so, she maximally entangles her qubits $A$ with the channel $U$ using $\magn{A}$ Bell pairs $\ket{\mathrm{EPR}}_A = \prod_{\magn{A}} (\ket{00}+\ket{11}) / \sqrt{2}$ as illustrated in Fig.~\ref{fig:fig3}a. Using local operations on her qubits $A$, Alice has complete control over the information entering the channel $U$; in particular, to send a quantum state $\ket{\psi}$ into the encoding circuit $U$, Alice need only project her maximally-entangled qubits $A$ onto the desired state $\ket{\psi}$.

Bob, another observer whose qubits $B$ are maximally-entangled with the remaining channel inputs, attempts to recover Alice's quantum information by collecting a subset $R$ of the output qubits. Of course if Bob has access to all of the output qubits $\magn{R} = N$ then his ability to reconstruct Alice's information is trivially guaranteed by the unitarity of $U$. Surprisingly, however, if the operator $U$ is strongly scrambling then Bob can reconstruct Alice's information using only a handful $\magn{R}_{\mathrm{min}} = \magn{A} + k$ of output qubits, with $k$ an $\mathcal{O}(1)$ constant independent of $\magn{R},\magn{A},N$. We can see why by examining the information content available to the regions $A,B,R,\overline{R}$.

The EPR pairs held by Alice and Bob in regions $A,B$ serve to convert the operator $U$ into a pure state $\ket{U}$ via the channel-state correspondence $\ket{U} = U \ket{\mathrm{EPR}}$ \cite{hosur2016chaos} as illustrated in Fig.~\ref{fig:fig3}a. We may therefore use entropy measures on the state $\ket{U}$ to quantify the amount of information that is accessible to various observers. In particular, the mutual information
\begin{align}
    \label{eq:scramblemutinfo}
    I_2(A:RB) &= S_A + S_{RB} - S_{ARB} \nonumber \\
        &= S_{A} + S_{A\overline{R}} -S_{\overline{R}} \nonumber \\
        &= 2 \magn{A} \ln 2 - I_2(A:\overline{R})
\end{align}
quantifies the amount of information shared between the region $A$ in Alice's control and the regions $R,B$ in Bob's (we have dropped the superscripts $S^{(2)},I_2^{(2)}$ for notational simplicity). In the last two lines we have used the fact that $S_{ARB} = S_{\overline{R}}$ and $S_{RB} = S_{A\overline{R}}$ which follow from the unitarity of $U$. Bob's ability to reconstruct Alice's quantum information is guaranteed in principle by maximal mutual information $I_2(A:RB) \approx 2 \magn{A} \ln 2$ between region $A$ and region $R \cup B$ \cite{hayden2007black,yoshida2017efficient,Yoshida2019Disentangling}. As shown in Eq.~\eqref{eq:scramblemutinfo}, this is equivalent to having vanishing mutual information $I_2(A:\overline{R})$ between $A$ and $\overline{R}$ such that the output qubits $\overline{R}$ alone reveal nothing about Alice's information. If this is the case then Bob can afford to ignore (or erase) the region $\overline{R}$ and still recover Alice's information, so long as he maintains control over regions $R,B$. In this language, the relation of scrambling to quantum error correction becomes particularly clear: vanishing mutual information $I_2(A:\overline{R})$ implies that Alice's quantum information is protected from all errors, up to and including erasure, acting on the qubits $\overline{R}$.




\section{VI. Details of Numerical Simulations}

\subsection{A. Page Scrambling of Polarized States in Different Basis}

\begin{figure}[h!]
    \includegraphics[width=0.5\textwidth]{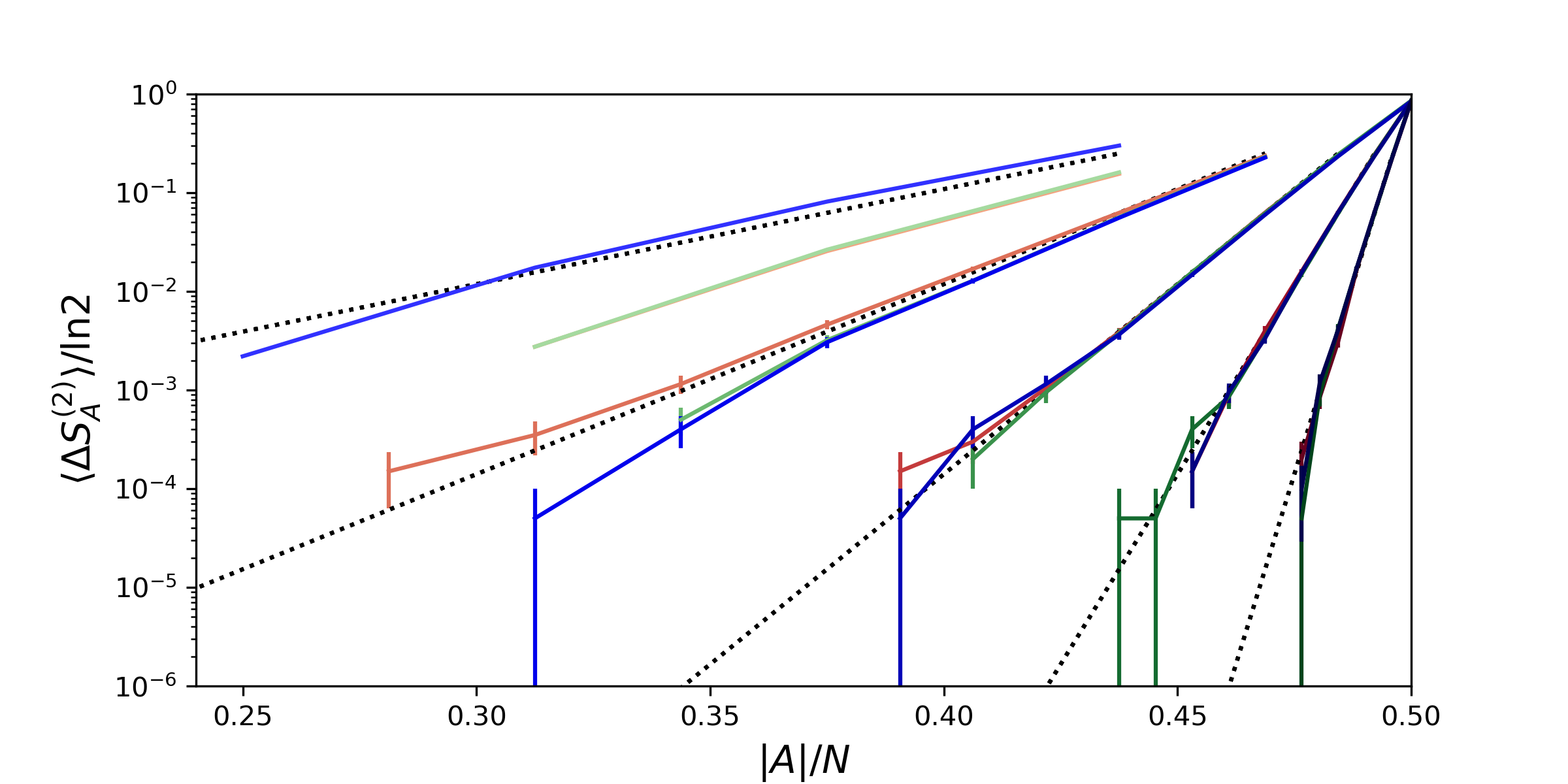}
    \caption{Average entropy deficits, $\langle \Delta \se_A \rangle / \ln 2$, 
    of the $x$ (red) ,$y$ (green), and $z$-polarized (blue) initial states for up to $2\mathrm{e}4$ 
    random subsystems $A$ compared to the Page limit (black dashed) computed from the Random Matrix Theory of $\mathrm{GF}(2)$
    for the system sizes $N=16,32,64,128,256$ (light to dark) at $t_*=2m=2\log_2(N)$.
    The points that are not shown for certain values of $|A|$, because 
    configurations with the entropy deficit of larger than $0$ could not be found due to the low occurrence of such configurations.
    }
    \label{SMfig:diffinitbasis}
\end{figure}

In this section, we show that the circuit $\chr$ can scramble a state regardless of the basis that is chosen for the initial state.
In Fig.~\ref{fig:fig2}, the initial state of $\ket{0}^{\otimes N}$, where $N$ is the system size, is computed 
to provide the various numerical evidence of the scrambling nature of the circuit.
Shown in Fig~\ref{SMfig:diffinitbasis} are the average entropy deficits of random subsystems $A$ of the 
$x$,$y$, and $z$-polarized initial states, which is the same quantity computed in Fig.~\ref{fig:fig2}c of the main text for the $z$-polarized initial state. As it can clearly be seen in the figure, the output states 
at the number of interaction layers $t=2m$ saturates the Page limit, showing that the circuit scrambles 
regardless of the choices of the basis of the initial states.

\subsection{B. Simulating Probabilistic Decoding with Quantum Trajectories}

While the Hayden-Preskill thought experiment tells under which conditions Bob can in principle recover the quantum state that has been sent by Alice into the scrambling circuit $\chr$, it does not tell how to do so in practice. One possibility to teleport a quantum state of Alice to Bob, is to add a probabilistic decoder \cite{yoshida2017efficient,Yoshida2019Disentangling} to the Hayden-Preskill thought experiment as shown in Fig.~\ref{fig:fig4}a of the main text. The decoder circuit $\chr^*$ is the complex conjugate of the scrambling circuit.

For the simulations presented in Fig.~\ref{fig:fig4}, we use a scrambling circuit having $N=8$ input qubits. Alice owns one qubit, referred to as $q_A$ in the following. As described in section \hyperref[sec:aydenpreskill]{V} of the supplemental material, Alice maximally entangles one input qubit of $\chr$ with her qubit. Bob maximally entangles the remaining input qubits of $\chr$ with qubits owned by him. The decoder circuit $\chr^*$ acts on these qubits owned by Bob and one additional qubit that is maximally entangled with a further qubit of Bob, referred to as $q_B$. The qubits' evolution under the scrambling and decoder circuits is calculated using the state vector simulator of the open-source framework Qiskit \cite{Qiskit}. After the calculation, we sample over all subsystems of size $|R|$, and evaluate the projector on the corresponding EPR pairs. The projection succeeds with the probability $P_{\EPR}$. In case of success, we evaluate the projector on the EPR pair between qubit $q_A$ and $q_B$. Such an EPR pair, which is measured with probability $F_{\EPR}$, is a resource for teleporting a quantum state of Alice to Bob.

To account for decoherence by an effective model, we add dephasing noise implemented via the depolarizing quantum error channel
\begin{equation}
D(\rho) = (1-p) \rho + p \frac{I}{2}
\end{equation}
to all qubits after each of the $2m=6$ interaction layers of the circuits. Here, $p$ is the error rate per qubit and per interaction layer. The error channel is treated with a quantum trajectory approach where we randomly sample over $6 \times 10^4$ realizations of the total circuit and average over the measurements.

For realizing the controlled-$Z$ gate within the simulations, we use the following scheme described in \cite{Jaksch2000}: We assume that $\ket{1}$ can be excited globally to the Rydberg state $\ket{r}$ by a $\pi$ pulse, using a laser whose Rabi frequency is much larger than the van der Waals interaction $V_\text{vdW}(R) \sim 1/R^6$ between adjacent atoms separated by the interatomic distance $R=R_{nn}$. For realizing the gate, we first transfer all the population from $\ket{1}$ to $\ket{r}$ by a global $\pi$ pulse. After waiting for the time $\pi / V_\text{vdW}(R_{nn})$, the state $\ket{rr}$ of two adjacent atoms has picked up the phase $\pi$ and we bring back the population to $\ket{1}$ by another global $\pi$ pulse.
We separate neighboring atoms by $2R_{nn}$ if we do not like to perform a controlled-$Z$ gate between them. Hereby, the picked up phase gets strongly suppressed by the fast decay of the van der Waals interaction.



\clearpage
\newpage



\end{document}